# Band offsets at amorphous-crystalline Al$_2$O$_3$–SrTiO$_3$ oxide interfaces

Dana Cohen-Azarzar, Maria Baskin, Lior Kornblum[a]

*The Andrew & Erna Viterbi Dept. of Electrical Engineering, Technion – Israel Institute of Technology, Haifa 32000 – Israel*

2D electron gases (2DEGs) formed at oxide interfaces provide a rich testbed for fundamental physics and device applications. While the discussion of the physical origins of this phenomenon continues, the recent discovery of oxide 2DEGs at non-epitaxial interfaces between amorphous and crystalline oxides provides useful insight onto this debate. Furthermore, using amorphous oxides offers a low-cost route towards realizing 2DEGs for device applications. In this work, the band offsets of a simple model system of amorphous-crystalline oxide interface are investigated. The model system consists of amorphous Al$_2$O$_3$ grown on single-crystalline (001) SrTiO$_3$. X-ray photoelectron spectroscopy is employed to study the chemical states, band gap and band offsets at the interface. The density of ionic defects near the interface is found to be below the detection limit, and the interface is found to be insulating. Analysis of the relative band structure yields significant interfacial barriers, exceeding 1.05 eV for holes and 2.0 eV for electrons. The barrier for holes is considerably larger than what is known for related materials systems, outlining the promise of using amorphous Al$_2$O$_3$ as an effective and simple insulator, an important building block for oxide-based field effect devices.

___________________________

[a] liork@technion.ac.il

## I. INTRODUCTION

The first observation of 2D electron gases (2DEGs) at an epitaxial interface between two insulating oxides[1] has quickly led to the discovery of rich and unexpected physics.[2–4] These findings have sparked tremendous interest in multiple communities, ranging from fundamental physics to materials science and to the engineering of various electronic and optoelectronic devices.[5–10]

The underlying physics of 2DEG formation has been narrowed down to two leading mechanisms: a polarization mechanism and an ionic defects mechanism. The polarization mechanism, or the *polar catastrophe*, argues that polar discontinuity at the interface, between a non-polar substrate, typically $SrTiO_3$, and a polar epitaxial overlayer such as $LaAlO_3$, results in a diverging potential across the latter. The potential divergence is mitigated by an electronic reconstruction that often occurs above a critical thickness. In this case, ½ an electron per unit cell is transferred from the top surface to the interface, where it occupies the empty Ti 3d-orbitals of $SrTiO_3$.[11] The recent observation of the long-sought 2D *hole* gas at this top interface[12] further supports the polarization mechanism.

Alternatively, the ionic mechanism highlights the role of defects as the cause of 2DEGs, typically defects that can be formed during the epitaxial growth of the overlayer[13] (e.g. $LaAlO_3$). Some possible defects can act as dopants near the surface of $SrTiO_3$. The most studied suspect in this context is the oxygen vacancy, a well-known electron donor in $SrTiO_3$. Oxygen vacancies were shown to form readily at typical growth conditions of $LaAlO_3$,[14] resulting in interface conduction.[15–18] Additional ionic suspects for the formation of 2DEGs are ions such as La, that can diffuse into $SrTiO_3$ during epitaxial growth, where they function as dopants,[19–22] as well as other electronically-active defects.[23,24] Ample experimental evidence supporting both mechanisms suggests that there is no single culprit behind the 2DEG phenomena.

To make things more interesting, it was later discovered[25,26] that 2DEGs can be formed at non-epitaxial interfaces, between a single-crystal substrate and amorphous oxides. Chemical evidence correlated the presence of 2DEGs and interfacial $SrTiO_3$ oxygen vacancies. This observation demonstrates that similar 2DEGs can be obtained in a material system where the polarization mechanism is not possible, due to the absence of long-range order. These observations do not categorically rule out the polarization mechanism, rather they underscore the complexity of the problem and highlight the importance of the surface chemistry, the oxidation states and interface electrostatics. A key aspect of this picture is the relative offsets between the bands at the interface. This aspect has been thoroughly studied in polar/non-polar *epitaxial* interfaces such as $GdTiO_3/SrTiO_3$,[27,28] $\gamma$-$Al_2O_3/SrTiO_3$,[29,30] $NdTiO_3/SrTiO_3$,[31,32] $LaNiO_3/SrTiO_3$,[27] $SmTiO_3/SrTiO_3$,[28] $LaCrO_3/SrTiO_3$[33] and $LaAlO_3/SrTiO_3$,[22,34–37] providing valuable physical insight, such as the existence[33] or absence[29,38] of polarization-induced internal fields, substrate band bending[29] and effects of ion intermixing[22] and oxygen vacancies.[36]

Despite this wealth of data on crystalline epitaxial interfaces, the band offsets at amorphous/crystalline oxide interfaces have not been addressed. These systems allow a case study of an electronic structure of interfaces in the absence of polar fields. Motivated by this knowledge gap, we study the surface chemistry, electronic structure and relative band offsets at an amorphous-crystalline interface between $Al_2O_3$ and $SrTiO_3$ (STO). This structure is further gaining recent attention as a tunneling junction in spintronic devices,[39,40] where better understanding of the electronic structure is expected to improve performance.

## II. EXPERIMENTAL

$TiO_2$ termination was performed on (001) undoped, 0.05 and 0.5%(wt) Nb-doped STO substrates (Shinkosha Ltd.) based on the 'extended Arkansas' method.[41] This process started with solvent sonication cleaning, followed by a 3:1 HCl-HNO$_3$ treatment and a two-step anneal, starting with 1,000°C for 1hr in air and completed with 650°C for 30 min in flowing $O_2$. Amorphous $Al_2O_3$ layers were grown by atomic layer deposition (ALD, Ultratech/Cambridge Nanotech Fiji G2) using trimethyl-aluminum (TMA) and water as the precursors at a substrate temperature of 300°C. 4 and 10 nm thick layers were grown: the thin layer puts both the $Al_2O_3$ and the STO substrate within the probing depth of x-ray photoelectron spectroscopy (XPS), whereas only $Al_2O_3$ is probed in the thick $Al_2O_3$ layer. The thicknesses of the layers were found to be in close agreement with the nominal values using x-ray reflectivity measutments. These layers are referred to as Thin $Al_2O_3$ and Thick $Al_2O_3$ henceforth.

An atomic force microscopy (AFM, Asylum MFP-3D Infinity) image was acquired in tapping mode from the surface of the Thick $Al_2O_3$ grown on an undoped STO substrate (Fig. 1). The observation of long range atomically-flat terraces with atomic height steps from the underlying STO substrate indicates that the roughness of the thin $Al_2O_3$ layer is well below the ~0.4 nm step height. No evidence of $Al_2O_3$ crystallinity was found in x-ray diffraction data acquired from the thick sample (Fig. S1, supplementary material). XPS (5600 Multi-Technique system, PHI) was acquired using monochromated Al Kα source (1486.6 eV) and a pass energy of 11.75 eV. Data was fit with the CasaXPS software using a Shirley background and a 30% Lorentzian-Gaussian ratio.

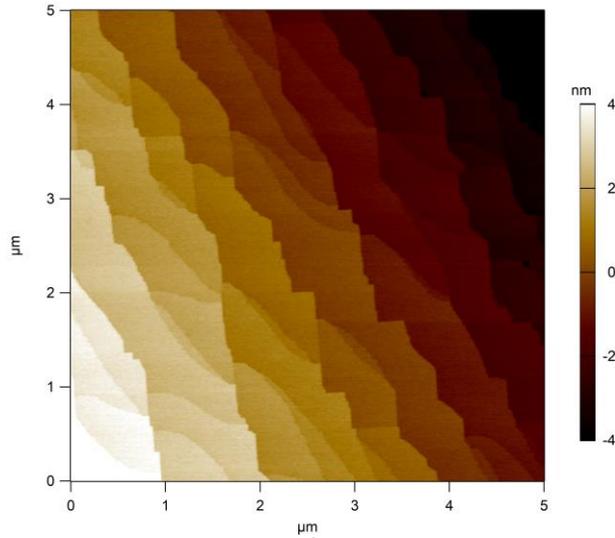

FIG. 1. Tapping mode AFM image of the surface of a 5×5 μm$^2$ region of the surface of a thick $Al_2O_3$ layer (10 nm) on undoped STO.

## III. RESULTS AND DISCUSSION

The Ti $2p_{3/2}$ XPS spectrum acquired from the thin $Al_2O_3$ and bare undoped STO samples (Fig. 2a) shows a well-behaved +4 state,[42] and no +3 oxidation states are observable at lower binding energies,[26,43] further validating the surface preparation procedure. The thick $Al_2O_3$ layer is used for studying the properties of $Al_2O_3$ without interference from the substrate. The Al 2p spectrum (Fig. 2b) shows well-behaved features that are fit with one doublet having a 0.4 eV separation,[44] showing a single oxidation state consistent with $Al_2O_3$.[45]

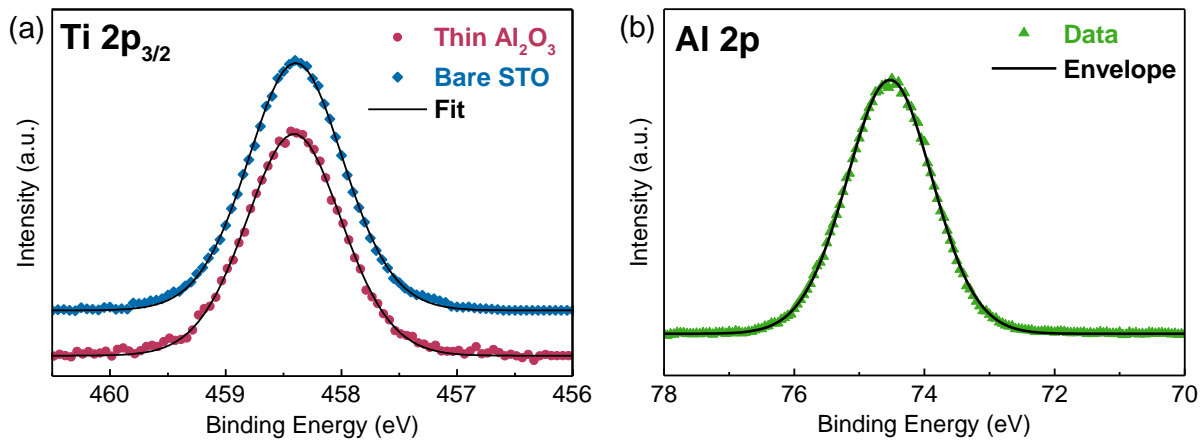

FIG. 2. (a) Ti $2p_{3/2}$ XPS spectra of the clean surface of a bare substrate (blue) superimposed on the Ti $2p_{3/2}$ spectrum, obtained from underneath the thin $Al_2O_3$ (4 nm) layer (red). (b) Al 2p spectrum of a Thick $Al_2O_3$ (10 nm).

Analysis of the O 1s region (Fig 3) reveals a major component that is ascribed to $Al_2O_3$ ('Peak 1'), and a minor moiety ('Peaks 2'), attributed to surface contamination. The uncertainty in the position of the small contamination peak has negligible effect on the position of the major O 1s component of the Al$_2$O$_3$ film.[46] The distance of the onset of the energy loss tail[47,48] from the major O 1s (Al$_2$O$_3$) peak yields a bandgap of 6.6±0.2 eV (denoted by a horizontal arrow in the inset of Fig. 3). This value is in agreement with previous reports for amorphous Al$_2$O$_3$.[44,48,49]

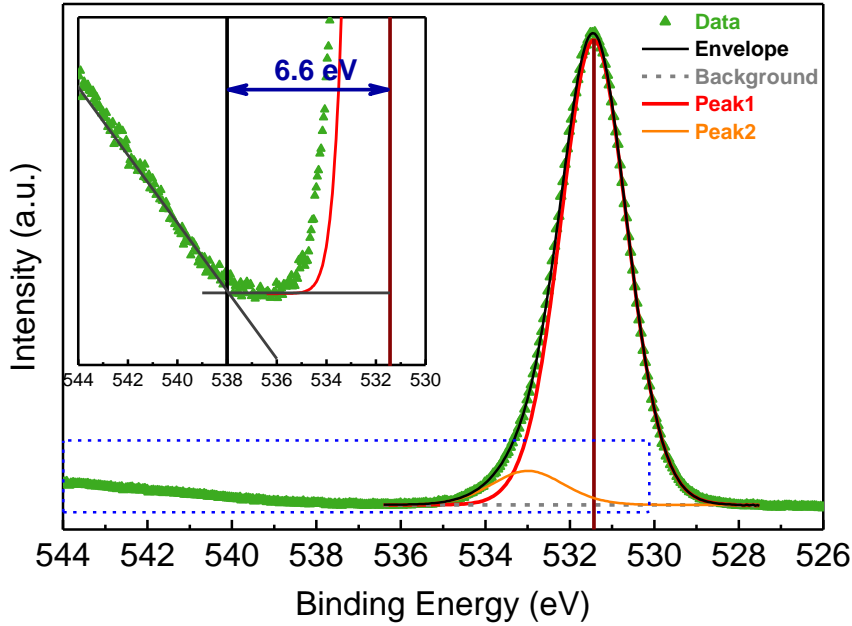

FIG. 3. O 1s spectrum of a Thick Al$_2$O$_3$ (10 nm). The inset shows a magnified region of the energy loss tail, taken from the dashed blue rectangle. The vertical lines represent the centroid of the major O 1s peak (brown) and the intersection of the linear fit of the loss tail with the background[48] (black). The distance between the lines (blue horizontal arrow) denotes the band gap of Al$_2$O$_3$.

Al contacts[50] were deposited on the corners of the undoped thin alumina on SrTiO$_3$ sample (e-beam deposition) after scratching it to contact the buried interface.[51] The resistance of the film was beyond the measurement limit (>5MΩ); we interpret this limit to indicate a sheet carrier density below $10^{11}$ cm$^{-2}$. This observation is in agreement with the absence of Ti$^{+3}$ signal in the Ti 2p$_{3/2}$ spectrum acquired from the Al$_2$O$_3$-STO interface (thin Al$_2$O$_3$), Fig. 2a. Similar observations were also made by Susaki et al. using hard x-ray photoemission spectroscopy (HAXPS) at polar LaAlO$_3$-STO interfaces.[37] Considering the background, we conclude that Ti$^{+3}$ could account for as much as 1% of the signal. A previous report of Al$_2$O$_3$-STO grown by ALD has shown a sheet carrier density of $3·10^{12}$ cm$^{-2}$ and corresponding Ti$^{+3}$ features, which were ascribed to surface reduction by the TMA precursor during the 300°C ALD process.[26] While the growth temperature employed here is similar, we attribute the absence of surface reduction here to variations in growth and control parameters between the different ALD systems. The ability to obtain an insulating interface without significant oxygen vacancies is useful for several applications; insulator-STO interfaces with low ionic defect densities are sought for various devices, and currently complex routes are employed to form such interfaces.[52,53]

Alignment of the energy scales of the different samples was done as follows: First, the energy scales of both the thick and thin Al$_2$O$_3$ layers were aligned so that the Al 2p$_{3/2}$ peak of each is positioned at 74.4 eV[44] (dashed vertical line in Fig. 4). The energy scale of the bare STO substrate was then aligned as such that the Ti $2p_{3/2}$ peak is at the same energy as the features acquired from underneath the thin $Al_2O_3$ layer (Fig. 4). This systematic alignment of the energy scale allows the direct comparison of the valence band edges[54–56] of $Al_2O_3$ and STO, yielding the valence band offset of 1.35±0.2 eV at the $Al_2O_3$-undoped STO interface. Kormondy and coworkers reported an offset of 0.9 eV for crystalline γ-$Al_2O_3$, epitaxially grown on STO,[30] and while γ-$Al_2O_3$ has a different band structure compared to the amorphous phase, we note the similarity of these values.

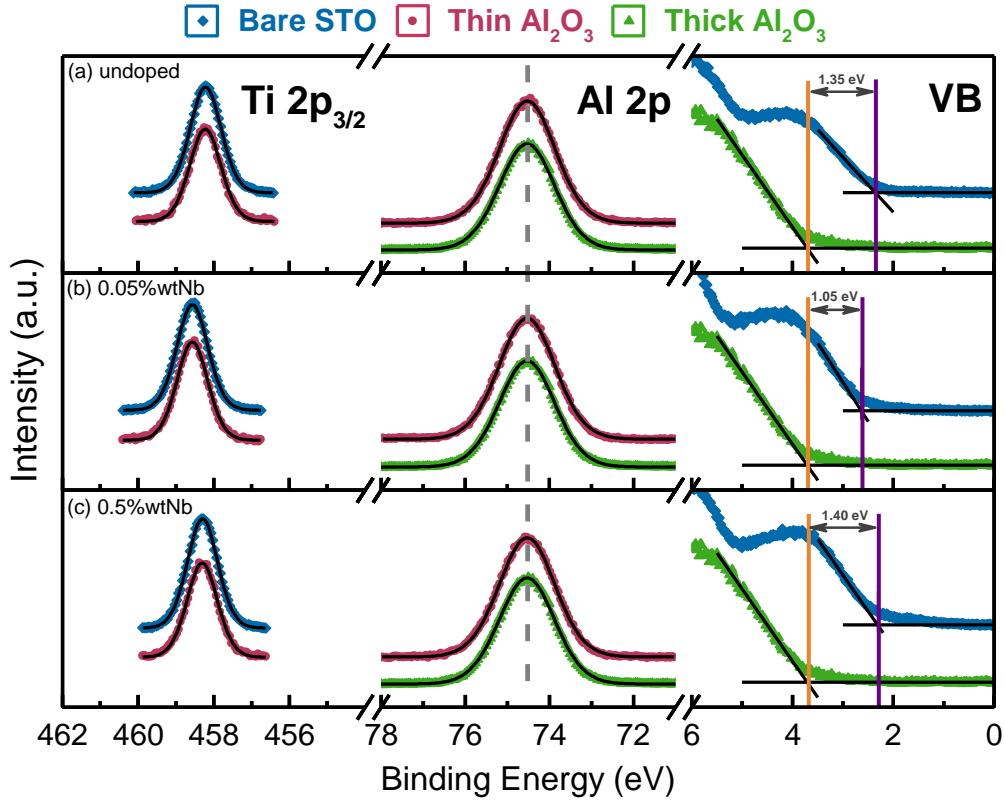

FIG. 4. Band alignment analysis of the $Al_2O_3$-STO interface. The Al 2p, Ti $2p_{3/2}$ and valence band (VB) spectra of a Bare STO substrate, Thin $Al_2O_3$ (4 nm) and Thick $Al_2O_3$ (10 nm) are shown for shown for (a) undoped, (b) 0.05%(wt) and (c) 0.5%(wt) Nb-doped STO substrates. The dashed grey line represents the energy alignment to an energy of 74.4eV. The Thick $Al_2O_3$ data (green triangles) is duplicated from panel a to panels b and c for clarity.

This analysis was further extended to study 0.05 and 0.5%(wt) Nb-doped STO substrates and resulted in valence band offsets of 1.05±0.2 and 1.4±0.2 eV, respectively (Fig. 4b and 4c, Table I). These valence band offsets are considerably larger than the reported values of 0.0-0.6 eV for the epitaxial $LaAlO_3$-STO interface.[34,36,37] Schütz et al. reported a ~0 eV valence band offset for epitaxial γ-$Al_2O_3$-STO, and a large ~3.8 eV conduction band offset.[29] The larger valence band offset measured here suggests that amorphous $Al_2O_3$ may be a better insulator than the more common $LaAlO_3$ and γ-$Al_2O_3$ for field effect devices.[10] This observation highlights the promise of ALD-grown $Al_2O_3$ to provide a simple route towards field effect devices based on oxide 2DEGs. We note that while insulating $AlO_x$ and $LaAlO_x$ have been previously used for lateral spacers or hard masks,[57–59] they have not been addressed or studied in the current context of barriers and energy alignment. Interestingly, Schütz et al. have further reported a ~0.6 eV downward band bending in STO,[29] whereas in the current work band bending is estimated to be negligible (Table S1 and discussion therein, Supplementary Material). The absence of band bending here highlights the role of the polarity of γ-$Al_2O_3$ in bending of the STO bands to compensate for this polar field.

Table I. Summary of the band offsets of $Al_2O_3$ with different STO substrates.

| Substrate | Valence band offset ±0.2 eV | Conduction band offset ±0.3 eV |
|---|---|---|
| Undoped STO | 1.35 | 2.05 |
| 0.05%(wt) Nb | 1.05 | 2.35 |
| 0.5%(wt) Nb | 1.4 | 2.0 |

Combining the measured valence band offsets with the bandgap of $Al_2O_3$ (6.6±0.3 eV, Fig. 3) and the bandgap of STO (3.2 eV, which does not change over the doping range used here[56]), the conduction band offsets are determined as 2.05, 2.35 and 2.0±0.3 eV for undoped, 0.05% and 0.5%(wt) Nb – $Al_2O_3$ interfaces, respectively (Fig. 5, Table I). An uncertainty range of ±0.2 eV is estimated from the determination of the valence band edge and the band gap (accumulating to ±0.3 eV when both are factored in), whereas the peaks were fit with uncertainties <0.05 eV. We therefore conclude that substrate doping, within the ranges studied here, has a small effect on the band alignment compared to these uncertainties. The slight deviation of the 0.05%(wt) sample from the others is possibly real, being outside the uncertainty range.

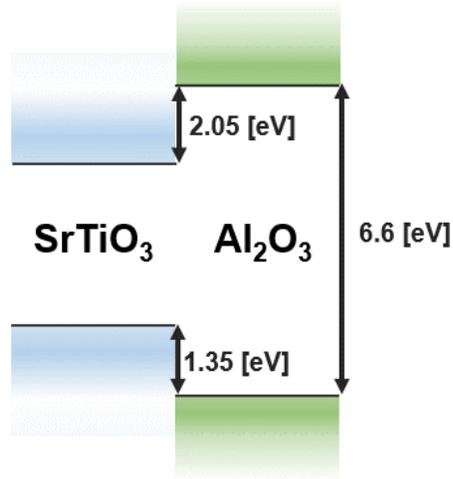

FIG. 5. Schematic relative band structure at the $Al_2O_3$-STO interface for the undoped substrate (Table I).

## IV. CONCLUSION

Non-polar amorphous $Al_2O_3$-STO oxide interfaces prepared by a simple, scalable, low-temperature process were studied. The interface with undoped STO was found to be insulating, with a low density of ionic defects – below the detection limit of XPS. The large band gap of amorphous $Al_2O_3$ results in offsets larger than those reported for related material systems, constituting significant interfacial barriers for both holes and electrons. The effect of STO doping on the band offsets is not significant, in the range of 0-0.5%(wt) Nb doping. These results highlight the potential of $Al_2O_3$ as an insulator for STO-based oxide electronics, constituting a promising building block for field effect devices.

## SUPPLEMENTARY MATERIAL

Online Supplementary Material includes microstructural analysis of $Al_2O_3$ with x-ray diffraction, and discussion of the band bending and built-in potentials, with a summary of the peak parameters from Figure 4.

## ACKNOWLEDGMENTS

This work was supported by the German-Israeli Foundation (GIF) for Scientific Research and Development Young Scientists Program (Contract I-2491-401) with partial support from the Technion's Russell Berrie Nanotechnology Institute (RBNI). Sample fabrication was done with the support of the Technion's Micro-Nano Fabrication unit (MNFU). The authors thank Dr. Larisa Burstein (Tel-Aviv University) for her assistance with XPS measurements, and Dr. Boris Meyler, Valentina Korchnoy and Tatiana Beker (Technion) for their valuable assistance with fabrication and processing of the samples. Lior Kornblum is a Chanin Fellow.

# Supplementary Material for Band offsets at amorphous-crystalline Al$_2$O$_3$–SrTiO$_3$ oxide interfaces

Dana Cohen-Azarzar, Maria Baskin, Lior Kornblum[a)]

The Andrew & Erna Viterbi Faculty of Electrical Engineering, Technion – Israel Institute of Technology,

Haifa 32000 – Israel

**Microstructural Analysis of Al$_2$O$_3$**

X-ray diffraction (XRD) measurements were conducted to rule out potential crystallinity of the Al$_2$O$_3$ layer, using a Rigaku Smartlab diffractometer with a 2-bounce Ge (220) monochromator. The diffraction pattern acquired for the thick (10 nm) Al$_2$O$_3$ film is presented in Fig. S1, only substrate peaks are observable, verifying the amorphous nature of the layer.

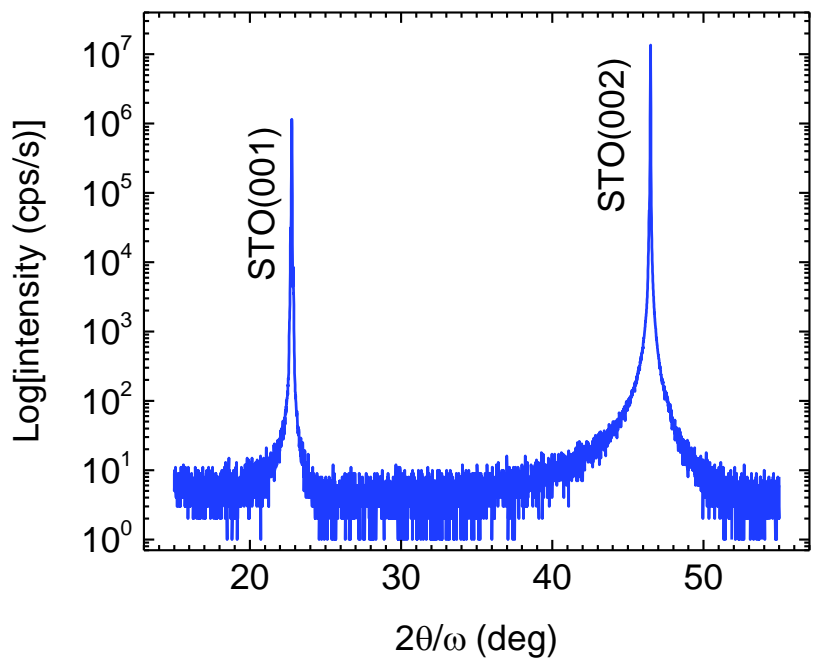

FIG. S1. X-ray diffraction data acquired from the thick Al$_2$O$_3$-STO structure.

**Discussion of Band Bending and Built-in Potentials**

The effect of STO band bending on the overall analysis was found to be negligible. Examining the three different 'bare' substrates, the Ti $2p_{3/2}$ to the VB maximum (VBM) energy difference was determined as: 455.87, 455.93 and 455.99 eV for undoped, 0.05% and 0.5%(wt) Nb-doped samples, respectively. This 0.12 eV difference is smaller than our conservative estimation of a ±0.2 eV uncertainty in the VBM determination. The fact that a metallic [0.5%(wt) Nb-doped] and an insulating (undoped) substrates show such small differences indicates that while band bending may exist, its effect on the interpretation of the data is negligible compared to the experimental uncertainty. Importantly, the full width at half maximum (FWHM) of the Ti $2p_{3/2}$ peak is 0.97, 0.97 and 0.94 eV for undoped, 0.05% and 0.5%(wt) Nb-doped samples, respectively (Table S1). If significant band bending existed, a distinct doping-dependent broadening of this feature was to be expected,[1] further supporting the negligible effect of possible band bending on the interpretation of the data.

**Table S1.** Summary of the fitting parameters of all the features used in the band offset analysis. BE, FWHM and VBM denote binding energy, full width at half maximum and valence band maximum, respectively. The Al 2p peak was fit with a doublet that includes an additional Al $2p_{1/2}$ component having the same FWHM, 1:2 area ratio and a 0.4 eV higher BE.

| Doping %(wt) Nb | Sample | Ti $2p_{3/2}$ | | Al $2p_{3/2}$ | | VBM |
|---|---|---|---|---|---|---|
| | | BE (eV) | FWHM (eV) | BE (eV) | FWHM (eV) | BE (eV) |
| 0.5% | Thick Al$_2$O$_3$ | - | - | 74.40 | 1.48 | 3.7 |
| | Thin Al$_2$O$_3$ | 458.30 | 0.95 | 74.40 | 1.48 | - |
| | Bare | 458.30 | 0.94 | - | - | 2.31 |
| 0.05% | Thin Al$_2$O$_3$ | 458.57 | 0.96 | 74.40 | 1.50 | - |
| | Bare | 458.57 | 0.97 | - | - | 2.64 |
| undoped | Thin Al$_2$O$_3$ | 458.22 | 0.97 | 74.40 | 1.49 | - |
| | Bare | 458.22 | 0.97 | - | - | 2.35 |

This discussion doesn't rule out possible a built-in potential across the Al$_2$O$_3$ layer. However, the FWHM of the Al $2p_{3/2}$ peak is identical, within 0.02 eV, for all samples containing Al$_2$O$_3$; these include the thick and the thin Al$_2$O$_3$ layers, the later both with doped and undoped substrates (Table S1). A built-in potential is expected to be manifested in broadening of these features,[1] and it remains unlikely that a thick layer would present the same internal field as a thin layer, and that an insulating and conductive substrates would result in the same screening and thus built-in potential in amorphous Al$_2$O$_3$. However, a rigorous conclusion regarding built-in potentials cannot be based on these observations alone. Unlike the band offset analysis which only relies on relative energy differences, a reliable estimation of the built-in potential would require the use of the absolute values of the binding energies. Reliable absolute values require meticulous and accurate charge compensation and spectrometer calibration.[2,3] Due to the highly insulating nature of Al$_2$O$_3$ and the fact that one of the substrates is insulating as well, we did not attempt to estimate the built-in potential owing to the possible errors stemming from charging of the sample. This charging has no effect on the band offset measurements reported in the paper, since no absolute energy values are needed there.